\documentclass[10pt,prl,aps,twocolumn,superscriptaddress,floatfix]{revtex4-1}
\usepackage{amsmath,amssymb} 
\usepackage{txfonts,graphicx,bm,color}
\def\iv{$I$-$V$ }
\def\dvdi{$dV/dI$ }

\begin{document}
\title{Shapiro steps in charge-density-wave states driven by ultrasound}
\author{Michiyasu Mori}
\email{Author to whom correspondence should be addressed: mori.michiyasu@jaea.go.jp}
\affiliation{Advanced Science Research Center, Japan Atomic Energy Agency, 
	Tokai, Ibaraki 317-1195, Japan}
\author{Sadamichi Maekawa}
\affiliation{Advanced Science Research Center, Japan Atomic Energy Agency, 
	Tokai, Ibaraki 317-1195, Japan}
\affiliation{RIKEN Center for Emergent Matter Science, Wako 351-0198, Japan}
\affiliation{Kavli Institute for Theoretical Sciences, University of Chinese 
	Academy of Sciences, Beijing 100049, China}
\begin{abstract}
We show that ultrasound can induce the Shapiro steps (SS) in the charge-density-wave (CDW) state. 
When ultrasound with frequency $\omega$ and a dc voltage are applied, the SS occur at the current $I$ $\propto$ $n\omega$ with integer $n$. 
Even and odd multiples of SS are represented by two couplings between the CDW and ultrasound. Although an ac voltage bias with frequency $\omega$ induces the SS at $I\propto n\omega$, 
the ultrasound bias enhances the odd multiples more strongly than the even ones. This is the difference between the ultrasound and the ac voltage. 
Since the SS cause abrupt peaks in the \dvdi, the extreme changes in the \iv curve will be applied to a very sensitive ultrasound detector.
\end{abstract}

\date{2023.1.25}

\maketitle
%========================================================
The charge-density wave (CDW) is a macroscopic quantum state with twice the Fermi wave vector (2$k_F$) induced by pairing of electrons and holes. 
The CDW is created by the electron-phonon interaction below the critical temperature of the Peierls transition $T_c$, which opens a semiconducting gap at a Fermi surface (FS) and increases resistivity~\cite{peierls55}. 
The CDW is pinned by impurities 
or in the case that a spatial period of the CDW matches with a lattice constant, i.e., commensurability pinning. 
When an applied electric field surpasses a certain threshold, the CDW is depinned and begins to slide through the lattice, increasing conductivity~\cite{froelich54,allender74,lee74,fukuyama76,fukuyama78,lee79}. 
Remarkable transport properties were observed in transition metal trichalcogenides such as NbSe$_3$ and TaS$_3$~\cite{monceau76,takoshima80,monceau80,monceu85,gruner85}. 
A current generated by the sliding of CDW above the threshold dc field contains not only a dc but also oscillating components known as narrow-band noise~\cite{fleming79,gruner81,monceau82,brown84,brown85,thorne86}. 
When a dc and an ac voltages are applied together, steps appear in an $I$-$V$ curve~\cite{fleming79,gruner81,monceau82,brown84,brown85,thorne86}. 
These steps are induced by phase locking of an external frequency with an internal one. 
This mechanism is same to the Shapiro steps (SS) in a Josephson junction of superconductors. 
The steps of an \iv curve in the CDW state shall be referred to as "Shapiro steps"(SS).

Ultrasound has rich variety of applications. 
Ultrasonography is used in hospitals for biomedical diagnostics to view our bodies or to eradicate cancers. 
A touch sensor using ultrasound is installed in some monitors.  
The ultrasound attenuation technique was used to investigate the symmetry of superconducting order parameters~\cite{tsuneto61,sigrist91}. 
On a surface of a ferromaget, a surface acoustic wave is used to control a spin current~\cite{maekawa76,kobayashi17,xu20}. 
Recently, electrical control of a surface acoustic wave toward quantum information technology has been realized~\cite{shao22}. 
In contrast to the electromagnetic waves, ultrasound does not radiate into free-space, allowing for coherent information processing and manipulation with a minimum cross-talk between devices and their environments~\cite{maldovan13}. 
The ultrasound is generated and detected by a piezoelectric material such as lithium niobate. 
Improved ultrasonic sensitivity will broaden its use in quantum information technologies. 
Recently, the SS by mechanical vibration have been reported~\cite{nikitin}.

\begin{figure}[t]
	\centering
	\includegraphics[width=0.45\textwidth]{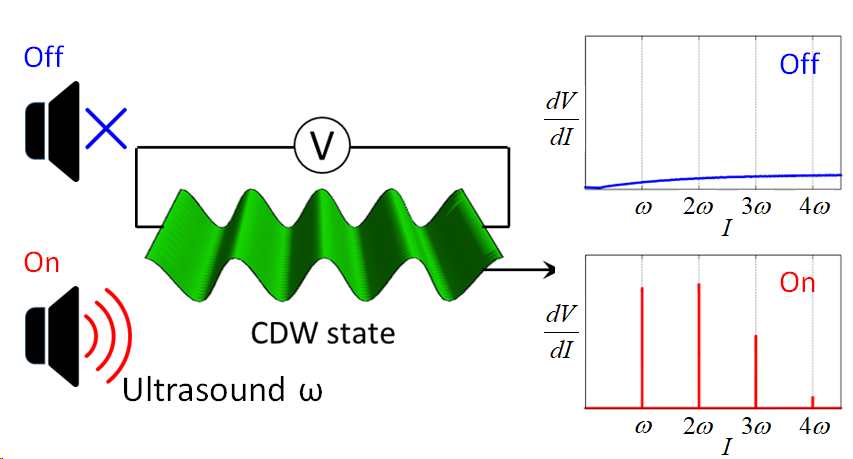}
	\caption{Schematic diagram of ultrasound detection using a CDW state. 
		The center picture depicts a voltage biasing the CDW state. 
		In the upper half, when the ultrasound is off, the \dvdi is monotonic as a function of $I$. 
		Peak structures develop at $\omega$ multiples in the bottom half when ultrasound with frequency $\omega$ is applied.\label{switching}}
\end{figure}

In this paper, we show that ultrasound can induce SS in the CDW state. 
When ultrasound with frequency $\omega$ and a dc voltage are applied, the SS occur at the current $I$ $\propto$ $n\omega$ with integer $n$. 
An ac voltage with frequency $\omega$ also induces the SS at $I\propto n\omega$. 
However, depending on the settings, the odd multiples of SS will be enhanced rather than the even ones due to two couplings of CDW to ultrasound; one contributes to even multiples, and the other to odd multiples.
This is the difference between the ultrasound and the ac voltage. Since the \dvdi shows sharp peaks due to the SS (See Fig. \ref{switching}), the drastic changes in the \iv curve will be applied to a highly sensitive ultrasound detector.

The charge density in a CDW state $\rho(x,t)$ is written by, 
\begin{equation}
	\rho(x,t)=\rho_0 + \rho_1 \cos\left[2k_Fx +\theta(x,t)\right], 
\end{equation} 
where averaged charge density $\rho_0$ and magnitude of $2k_F$-oscillation $\rho_1$ are constant. 
The dynamical behavior at low energies is given by, 
\begin{equation}
\mathcal{L}_{\rm 0} 
	=
	\frac{1}{4\pi v'}\!\!\int\!\! dx
		\left\{
			\bigl[\partial_t\theta(x,t)\bigr]^2 
		- v^2\big[\partial_x\theta(x,t)\big]^2
		\right\},\label{fukuyama}
\end{equation}
where $v'$ $\equiv$ $v^2/v_F$ with Fermi velocity $v_F$ and $v$ is the velocity of Goldstone mode in the CDW state~\cite{fukuyama76}. 
When dealing with one-dimensional electron systems, $\theta(x,t)$ is related to fermionic operators $\psi_{s\pm}$ of electrons using the bosonization given by, 
$\psi_{s\pm}$ $\sim$ 
$\exp(\pm i k_F x \pm \theta_{s\pm})$ so as to $\theta(x,t)$ $=$ $\sum_s\theta_{s+}+\theta_{s-}$ with spin $s$~\cite{tomonaga50,luttinger63,solyom,haldane81}. 
According to these equations, $\theta$ denotes the fluctuation of electron density $\delta n_{\rm e}$ around a FS. 
Considering $2k_F$ $=$ $2(2\pi/L)(N/2)$ with system size $L$ and number of electrons $N$,
$k_F$ corresponds to the electron density $N/L$ $\equiv$ $n_{\rm e}$. 
Then, $(1/\pi)[\partial \theta(x,t)/\partial x]$ is associated with $\delta n_{\rm e}$, because of 
the spatial derivative of the phase factors in $\psi_{s\pm}$ leads to the electron density, i.e.,  $\partial(2k_F x + \theta(x,t))/\partial x = 2k_F + \partial \theta(x,t)/\partial x =\pi(n_{\rm e}+\delta n_{\rm e})$~\cite{tomonaga50,luttinger63,solyom,haldane81}. 
The continuity equation $\partial_t n+\partial_x j=0$ defines the current $j$ $=$ $-(1/\pi)[\partial \theta(x,t)/\partial t]$. 

The key to transport characteristics is CDW pinning. We shall now look at a single impurity model. 
An impurity potential at $x$ = 0 is given by,
\begin{align}
	U_{\rm imp} 
		&= -u_0\rho_1 \!\!\int\!\! dx \cos \left[Q x+ \theta(x,t)\right] \delta \left( x\right),\\
		&= -u_0\rho_1 \cos \theta,\label{imp}
\end{align}
where $Q$ $\equiv$ $2k_F$, $\theta$ $\equiv$ $\theta(x=0,t)$, $u_0$ $>$ 0 and $\rho_1$ $>$ 0.
When ultrasound or surface acoustic wave is applied, the impurity is moved from its initial position
and the energy provided by Eq.~(\ref{snd}) is added,
\begin{align}
    U_{\rm snd} &= u_0 \rho_1 \!\!\int\!\! dx \cos \left[ Qx + \theta(x,t)\right]
	 \left[ {\delta \left( {x \!-\! \xi(t) } \right) - \delta \left( x \right)} \right],\\
	 &\sim -u_0\rho_1 \sin \theta \cdot \left[Q + \partial\theta/\partial x\right]\cdot\xi(t),\label{snd}
\end{align}
where  
$\partial\theta/\partial x \equiv \partial\theta(x,t)/\partial x |_{x\rightarrow \pm R_0}$~~\cite{nagaoka}. 
At the impurity site, $x$ = $0$, the spatial variation of $\theta(x,t)$ is singular and its derivative is not
continuous~\cite{teranishi79}. 
The right (left) derivative of $\theta(x,t)$ at the impurity site will be negative (positive) so as to fix the sign of $\partial\theta/\partial x \cdot \xi(t)$. 
For $Q\cdot\xi(t) \equiv \eta \cos (\omega t)$, the pinning potential and  the coupling to ultrasound are given by,
\begin{align}
	U&=  U_{\rm imp}+U_{\rm snd},\\
	&= -u_0\rho_1 \left[ \cos \theta + \sin \theta \cdot d(t) \right],\label{attenuation}\\
	d(t)&\equiv \eta\left( \chi\cos \omega t + \lambda \,\left|\cos \omega t \right|\right), \label{external}
\end{align}
where $\chi$ = 1, and $\lambda \equiv |\partial\theta/\partial x| /Q$ $>$ 0. 
In Eq. (\ref{external}), the first term will be dominant in most 
circumstances because $\lambda$ relates to $\delta n_e$ as $\lambda=|\partial\theta/\partial x| /Q=\delta n_{\rm e}/n_{\rm e}$.
It is usually less than 1, i.e., $\lambda < 1$. 
Below, $\eta$ = 1 is imposed for brevity and the following results do not change qualitatively. The magnitude of the oscillation at the impurity site is proportional to $\eta$, which can be regulated by the ultrasonic input power.  

The single impurity model can account for several important aspects of CDW transport properties. 
This is the limit of strong pinning.  
When pinning is weak enough that a single impurity cannot pin the phase of the CDW, the pinning potential is determined by balancing the increase in kinetic energy owing to charge density change
i.e., the second term of Eq. (\ref{fukuyama}), and an energy gain from phase-dependent impurity pinning energy~\cite{fukuyama78,lee79}. 
These are characterized by the pinning frequency in the optical conductivity~\cite{fukuyama78,lee79}. 
The spatial distortions of the phase in the ground state caused by the random distribution of impurities are an important characteristic of impurity pinning. If
the quantity of interest is not sensitive to such spatial 
fluctuation, the pinning will be 
characterized by an averaged potential proportional to $\cos\theta(x,t)$, where $\theta(x,t)$ is an averaged phase 
within an impurity distribution-determined domain~\cite{fukuyama85}. Hence, the above discussion will be valid even in a weak pinning region. 

Since electrons move together with lattice distortion, the effective mass of CDW is large and the dynamics of phase motion can be treated classically.
Similar to the SS in a Josephson junction of superconductors, the transport properties of CDW can be calculated by, 
\begin{equation}
	\alpha \frac{d^2\theta}{d t^2} + \beta\frac{d\theta}{d t}+U'(\theta)=F(t),\label{accel}
\end{equation}
with $U'(\theta)=dU/d\theta$. 
Equation (\ref{accel}) is interpreted as the acceleration $\alpha$$d^2\theta/d t^2$ by the effective driving force,
$F_{eff}=F(t)-U'(\theta)$, with the friction term $\beta$$d\theta/d t$. 
In an overdamped region, $\alpha\sim0$, the first term in Eq. (\ref{accel}) is neglected.  
When ultrasound is used to drive a CDW, 
the transport properties are determined by, 
\begin{equation}
	\beta\frac{d\theta}{d t} + u_0\rho_1 \left[ \sin \theta + \cos\theta\cdot d(t)\right]=f_{dc},\label{sound}
\end{equation}
where $f_{dc}$ corresponds to a dc voltage. 
Below, we will use the following parametrization, 
\begin{align}
	\frac{d\theta}{d t}+U'(\theta)&=V,\label{solve}\\
	U'(\theta)&=A\left[ \sin \theta + \cos\theta \cdot d(t) \right].\label{pinning}
\end{align}
The current $I$ $\equiv$ $\pi j$ is calculated by averaging of $\partial \theta(t)/\partial t$ in time as, 
\begin{align}
	I&=-\frac{1}{T}\int_0^T \!dt \left(\frac{\partial \theta(t)}{\partial t}\right)\label{current}.
\end{align}

\begin{figure}[h]
	\centering
	\includegraphics[width=0.45\textwidth]{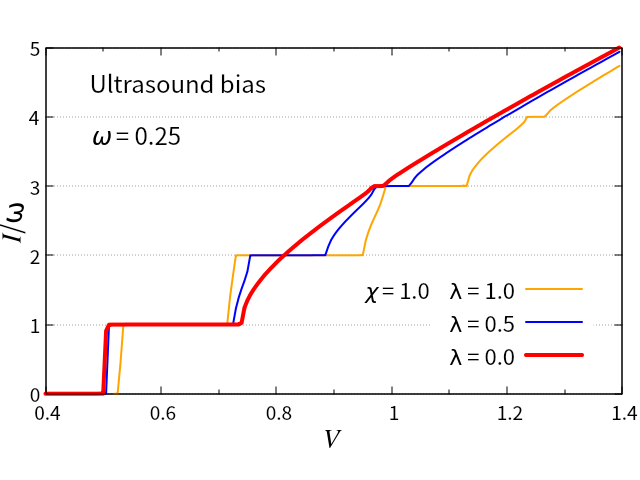}
	\caption{The $I$-$V$ curve driven by ultrasound with $\omega$ = 0.25, $A$ = 0.5, $\chi$ = 1.0, and $\lambda$ = 1.0 (orange), 0.5 (blue), and 0.0 (red).  
	The SS appear at $I/\omega$ = $n$ with integer $n$ except for $\lambda$ = 0.0, where  only odd multiples appear.\label{sound1}}
\end{figure}

\begin{figure}[h]
	\centering
	\includegraphics[width=0.45\textwidth]{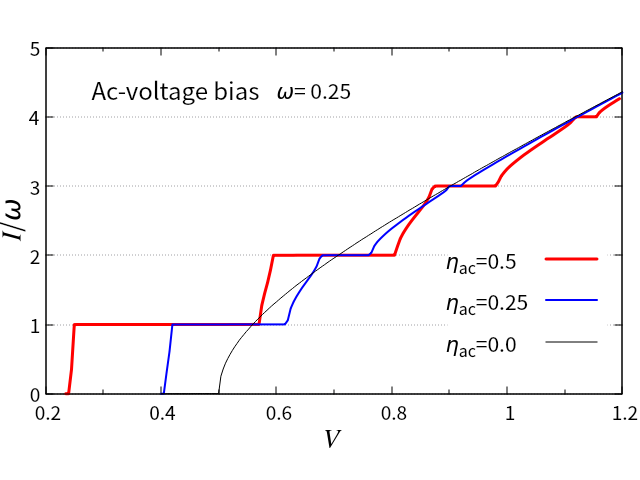}
	\caption{The $I$-$V$ curve driven by an ac voltage with $\omega$ =0.25, $A$ = 0.5, and $\eta_{\rm ac}$ =0.5 (0.25) for the red (blue) curve. The black curve is without ac voltage. 
		The steps appear at $I/\omega$ = $n$ with integer $n$. 
		\label{voltage}}
\end{figure}

Solving Eqs.~(\ref{solve}) and (\ref{current}), the $I$-$V$ curves are obtained as shown in 
Fig. \ref{sound1} with 
$\omega$ = 0.25, $\chi$ = 1.0 and $\lambda$ = 1.0 (orange), 0.5 (blue), and 0.0 (red).
The SS appear at $I/\omega$ = $n$ for $\lambda$ = 1.0 and 0.5. For $\lambda$ = 0.0, however, the SS appear at only odd multiples, i.e., $I/\omega$ = 2$n$+1, and even multiples are suppressed. This is a critical distinction between ultrasound and an ac voltage. When the system is driven by the ac voltage, 
the equation of motion is given by~\cite{gruner81,monceau82}, 
\begin{align}
	\frac{d\theta}{d t}&=V-A\sin \theta - \eta_{\rm ac} \cos (\omega t).\label{solve2}
\end{align}
Solving Eq.~(\ref{solve2}), the $I$-$V$ curves are obtained as shown in Fig.~\ref{voltage} with 
$\omega$ = 0.25 and $\eta_{\rm ac}$ = 0.5 (red), 0.25 (blue), and 0.0 (black). 
The SS appear at $I/\omega$ = $n$ similar to a Josephson junction of superconductors. 
In the case of Eq. (\ref{solve2}), assuming $\theta=at-b\sin(\omega t)$ with constants $a$ and $b$, and using the relation, $e^{iz\sin\theta}=\sum_n J_n(z)e^{in\theta}$, with Bessel function $J_n(z)$, 
the potential term is transformed as, $\sin\theta={\rm Im}\left[\sum_n J_n(b) e^{i(a-n\omega)t}\right]$. 
When $a=n\omega$, 
the time average of $\sin\theta$ is time-independent and leads to the SS. 
Since another potential term, such as $\sin\theta\cos\omega t$, is multiplied by the oscillation in the ultrasound bias Eq.~(\ref{pinning}), analytical estimation of the SS is problematic. 
However, using the guess, $\theta=at-b\cos(\sin\omega t)$, 
the potential term is written as, $\sin\theta={\rm Im}\left[\sum_{n>0,m}J_n(b)J_m(n)(1+(-1)^m)e^{i(a-m\omega)t-in\pi/2}\right]$. 
Then, the SS appear at $a=m\omega$ with even multiples. When $\lambda$=0, the oscillating term in Eq. (\ref{pinning}) is proportional to $\cos(\theta-\omega t)+\cos(\theta+\omega t)$. 
Hence, another guess of $\theta\pm \omega t =at-b\cos(\sin\omega t)$ could be possible, and the SS appear at odd multiples as shown in Fig. \ref{sound1}. 

The black line in Fig.~\ref{voltage} represents the threshold $V=A$, at which the solution changes its behavior. 
At $\eta_{\rm ac}$ = 0, Eq.~(\ref{solve2}) is analytically solved as,
$V\tan(\theta(t)/2)=A+(V^2-A^2)^{1/2} \tan[(V^2-A^2)^{1/2}t/2]$.
For $V < A$, the solution is monotonously merging to a constant, whereas, for $V > A$, the solution oscillatory increases or decreases with time. Considering Eq. (\ref{current}), the increase or decrease of $\theta$ in time leads to a finite value of $I$.  
Because of the phase-locking, the solution at the SS do not change its gradient.

\begin{figure}[h]
	\centering
	\includegraphics[width=0.45\textwidth]{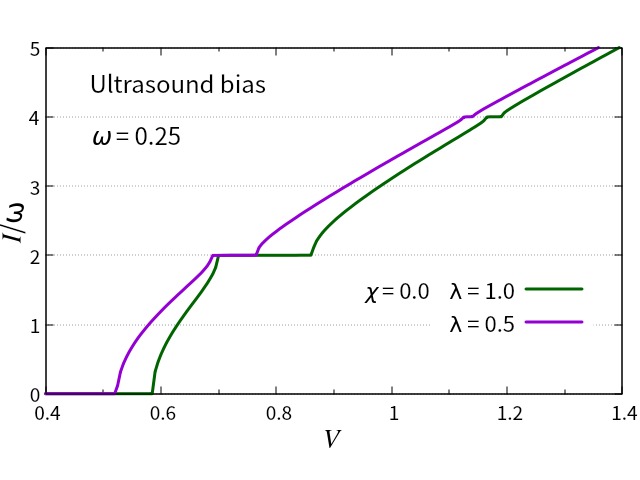}
	\caption{The $I$-$V$ curve driven by ultrasound for $\chi$ = 0.0 with $\omega$ = 0.25, $A$ = 0.5, and $\eta$ = 1.0 (0.5) for the green (purple) curve.  The SS appear at $I/\omega$ = 2$n$ with integer $n$.\label{sound2}}
\end{figure}

In Fig.~\ref{sound1}, without the $\lambda$-term in Eq.~(\ref{external}), the SS appear only at odd multiples. However, only with the $\lambda$-term, i.e., $\chi$ = 0.0, the \iv curves are presented in Fig.~\ref{sound2} for $\lambda$ = 1.0 (green) and 0.5 (purple).
Notably, the SS appear at even multiples, i.e. $I/\omega$ = 2$n$, which is in sharp contrast to Fig.~\ref{sound1}.  
The $\lambda$-term in Eq.~(\ref{external}), $\cos\theta|\cos \omega t|$, is not equivalent to $\cos(\theta-\omega t)+\cos(\theta+\omega t)$ due to $|\cos \omega t|$. Hence, $\theta=at-b\cos(\sin\omega t)$ will be one of the possibilities of even multiples. Another reason is that the external force proportional to $|\cos \omega t|$ is nearly equivalent to the 2$\omega$-oscillation.

We discovered that ultrasound 
can drive the CDW state in two ways, namely
$\chi$- and $\lambda$-terms in Eq.~(\ref{external}), which cause odd and even multiples in the SS, respectively.  
By definition $\chi$=1 and $\lambda=|\partial\theta/\partial x| /Q=\delta n_{\rm e}/n_{\rm e} < 1$, 
the odd multiples due to the $\chi$-term will be dominant in the \iv curves. 
This will help to distinguish between an ultrasound contribution and an electromagnetic one. 
For example, the red curve in Fig.~\ref{sound1} is driven by 0.5$\cdot\cos\theta\cdot\cos\omega t$ and another red curve in Fig.~\ref{voltage} is driven by 0.5$\cdot\cos\omega t$. 
Just a factor of $\cos\theta$ leads to completely different results. 
When ultrasound changes the dielectric constant of a material, an electromagnetic field would be simultaneously applied together with the ultrasound. In this case, the SS appear all of the integer multiples as $I/\omega$ = $n$. 
The step-width at even multiples will be suppressed, 
if the produced electromagnetic field weakens.
The electromagnetic bias is proportional to $\cos\omega t$, while the ultrasound bias has an additional factor $\cos\theta$ such as, $\cos\theta\cdot\cos\omega t$ and $\cos\theta\cdot|\cos\omega t|$. Even though the difference is just a factor of $\cos\theta$, the step-widths are determined by their functional shapes. In the case of electromagnetic bias, all step-widths decrease with $\eta_{\rm ac}$ as shown in Fig.~\ref{voltage}. There is not much of a distinction between even and odd multiples.
\begin{figure}[h]
	\centering
	\includegraphics[width=0.45\textwidth]{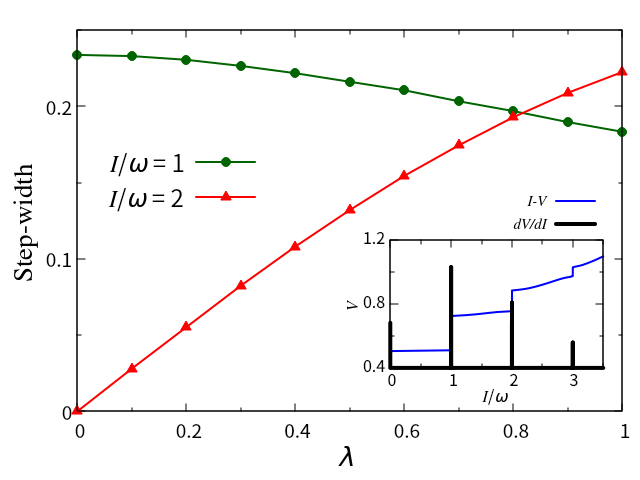}
	\caption{The $\lambda$-dependence of the step-width at $I/\omega$ = 1 (green) and 2 (red) with $\omega$ =0.25, $A$ = 0.5, and $\chi$ = 1.0. The inset shows the \iv curve (blue), which is the same as that in Fig. \ref{sound1}, and its derivative \dvdi (black). The peaks corresponding to the steps in the \iv curve appear at $I/\omega$= $n$ with integer $n$. 
	}\label{width}
\end{figure}
In Fig. \ref{width}, on the other hand, the step-width at $I/\omega$=1 and 2 are plotted as a function of $\lambda$ for $\chi$=1.0. The step-width at $I/\omega$=1 (green) drops modestly with $\lambda$, whereas the step-width at $I/\omega$=2 (red) climbs fast from zero. Since $\lambda$ is proportional to the density fluctuation, $\lambda$ will be less than 1, i.e., $\lambda=\delta n_{\rm e}/n_{\rm e} < 1$, and hence the odd multiples such as $I/\omega$=1 will be observed rather than even ones.
The inset of Fig.~\ref{width} shows the \iv curve (blue) and its derivative \dvdi (black) for $\lambda$ = 0.5. The peaks corresponding to the SS in the \iv curve appear at $I/\omega$ = $n$ with integer $n$. 
In this scenario, 
the \dvdi is measured around a bias voltage, such as $V$ = 0.8. 
We will see the sharp peak in \dvdi at $I/\omega$ = 1, only when ultrasound is applied. 
Furthermore, analogous 
to the Josephson junction of superconductors,
this transition will be quite rapid and abrupt. As a result, 
the SS in the CDW state will detect ultrasounds with high sensitivity. 

\begin{figure}[h]
	\centering
	\includegraphics[width=0.45\textwidth]{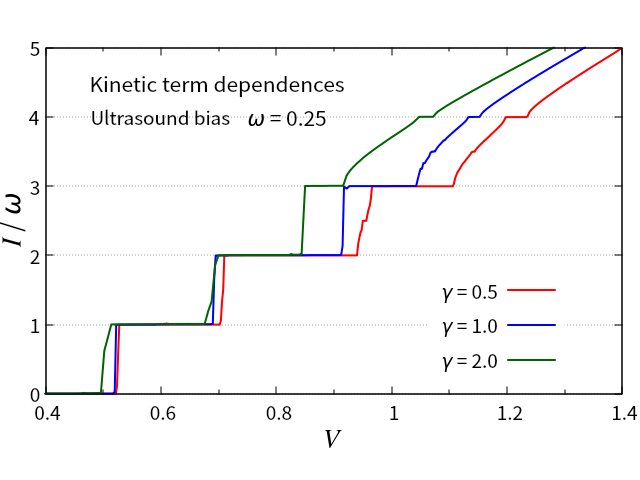}
	\caption{The $I$-$V$ curve driven by ultrasound with $\omega$ =0.25, $A$ = 0.5, 
			$\chi=\lambda$ =1.0, and 
			$\gamma$ = 0.5 (red), 1.0 (blue), and 2.0 (green). \label{kin}}
\end{figure}

We have so far discussed the SS in the overdamped region. 
To test the effect of the kinetic term, i.e., the first term in Eq. (\ref{accel}),  
the \iv curve is calculated using the following equation,
\begin{align}
	\gamma\frac{d^2\theta}{d t^2}+\frac{d\theta}{d t}+U'(\theta)&=V.\label{kinterm} 
\end{align}
The results are shown in Fig. \ref{kin} for $\omega$ =0.25, $A$ = 0.5, 
$\chi=\lambda$ =1.0, and 
$\gamma$ = 0.5 (red), 1.0 (blue), and 2.0 (green).
The SS arise independently of the kinetic term at $I/\omega$ = $n$. 
As a result, even if the kinetic term exists, the SS in the CDW state will respond sensitively to the ultrasound. 
In the cases of $\gamma$ =0.5 and 1.0, the \iv curves show tiny sub-harmonics at $I/\omega$=2.5 and 3.5, respectively. 
The subharmonic steps ($I/\omega$ = $n/m$) cannot be explained by the overdamped model with a single impurity, 
whereas the kinetic term changes the matching condition of the phase locking and can induce the subharmonic steps~\cite{ya84}. 
Some materials exhibit subharmonic steps, 
the amount of which may 
be explained by factors other than the kinetic term~\cite{brown84,brown85}.
Another possibility to reproduce the subharmonic steps is multiple impurities in the overdamped model~\cite{matsukawa87,matsukawa87jjap,matsukawa87jpsj2}. 
In this situation, the subharmonic steps 
become evident in the same way that 
the harmonic steps do.
	Bardeen, on the other hand, proposed a non-sinusoidal potential
	in which an averaged impurity distribution is periodic and the CDW phase tunnels between two stable states~\cite{thorne86,bardeen85,bardeen85book}. 
	The multiple impurities and the non-sinusoidal potential are outside our scope of this paper. Those will be studied in the near future.

We have shown that ultrasound can induce the SS in the CDW state. 
When ultrasound with frequency $\omega$ and a dc voltage are applied, the SS occur at the current $I$ $\propto$ $n\omega$ with integer $n$. 
Two couplings between the CDW and ultrasound constitute even and odd multiples of SS, respectively. Although an ac voltage with frequency $\omega$ induces the SS at $I\propto n\omega$, the ultrasound enhances the odd multiples more strongly than the even ones. This is the difference between the ultrasound and the ac voltage.
Since the \dvdi shows sharp peaks due to the SS (See Fig. \ref{switching}), the drastic changes in the \iv curve will be applied to a highly sensitive detector of ultrasound. 

%%%======================================================
\begin{acknowledgements}
We would like to thank Prof Niimi and Mr. Fujiwara for valuable discussions about the experiments on SAW in 1D CDW materials. Thanks are also to Prof Matsukawa for his important suggestions and valuable discussions. This work was supported by JSPS Grant Nos.~JP20K03810 and JP21H04987, and the inter-university cooperative research program (No.~202012-CNKXX-0008) of the Center of Neutron Science for Advanced Materials, Institute for Materials Research, Tohoku University. A part of the computations was performed on supercomputers at the Japan Atomic Energy Agency. 
S.M. is supported by JST CREST Grant (Nos. JPMJCR19J4, JPMJCR1874, and JPMJCR20C1)
and JSPS KAKENHI (nos. 17H02927 and 20H01865) from MEXT, Japan.
\end{acknowledgements}

\end{document}